%% file: socg_with_appendix_2019.tex
\documentclass[nolineno,a4paper,UKenglish]{socg-lipics-v2019}

\usepackage{epsfig}
\usepackage{graphicx}
\usepackage{color}
\usepackage{times}
\usepackage{amsthm}
\usepackage{amsmath}
\usepackage{amssymb}
\usepackage{xspace}
\usepackage{pb-diagram}

\makeatletter
\long\def\@makecaption#1#2{%
  \vskip\abovecaptionskip
  \sbox\@tempboxa{\small #1: #2}%
  \ifdim \wd\@tempboxa >\hsize
    \small #1: #2\par
  \else
    \global \@minipagefalse
    \hb@xt@\hsize{\hfil\box\@tempboxa\hfil}%
  \fi
  \vskip\belowcaptionskip}
\makeatother

\renewcommand{\paragraph}[1]{\vspace{1ex}\noindent{\textbf{#1}}}
\renewcommand{\subparagraph}[1]{\vspace{1ex}\noindent{\textit{#1}}}

\newcommand{\myproof}{\noindent{\sc Proof.~}}
\newcommand{\myeop}{\hfill\usebox{\smallProofsym}\bigskip}  %
\newsavebox{\smallProofsym}                            
\savebox{\smallProofsym}                               %
{
\begin{picture}(6,6)                                   %
\put(0,0){\framebox(6,6){}}                            %
\put(0,2){\framebox(4,4){}}                            %
\end{picture}                                          %
}

\newcommand {\mm}[1] {\ifmmode{#1}\else{\mbox{\(#1\)}}\fi}

\newcommand {\scalprod}[2] {{\langle #1 , #2 \rangle}}
\newcommand {\denselist}{\itemsep 0pt\parsep=1pt\partopsep 0pt}

\newcommand{\Fgroup}        {\mm{\sf{F}}}
\newcommand{\Ggroup}        {\mm{\sf{G}}}
\newcommand{\Hgroup}        {\mm{\sf{H}}}

\newcommand{\Hspace}        {\mm{\mathbb{H}}}
\newcommand{\Rspace}        {\mm{\mathbb{R}}}
\newcommand{\Sspace}        {\mm{\mathbb{S}}}
\newcommand{\Xspace}        {\mm{\mathbb{X}}}
\newcommand{\Domain}        {\mm{\Omega}}
\newcommand{\DFunction}[1]  {\mm{\cal D}_{{#1}}}
\newcommand{\RFunction}[1]  {\mm{\cal R}_{{#1}}}
\newcommand{\Power}[1]      {\mm{\cal P}_{0}{({#1})}}
\newcommand{\Length}[1]     {\mm{\rm Length}{({#1})}}
\newcommand{\One}           {\mm{{\bf 1}}}
\newcommand{\Hausdorff}[3]  {\mm{\cal H}_{#1}{({#2},{#3})}}
\newcommand{\dHausdorff}[3] {\mm{\cal H}_{#1}{({#2}\|{#3})}}

\newcommand{\JSLs}[2]       {\mm{{\rm JS}{({#1},{#2})}}}
\newcommand{\JS}            {\mm{{\rm JS}}}
\newcommand{\BRLs}[3]       {\mm{{\rm BR}_{#1}{({#2},{#3})}}}

\newcommand{\dist}          {\mm{\theta}}
\newcommand{\distj}         {\mm{\jmath}}
\newcommand{\disti}         {\mm{\imath}}
\newcommand{\distFn}[2]     {\mm{\theta}{({#1},{#2})}}
\newcommand{\distjFn}[2]    {\mm{\jmath}{({#1},{#2})}}
\newcommand{\distiFn}[2]    {\mm{\imath}{({#1},{#2})}}

\newcommand{\Bdist}[3]      {\mm{{D}_{#1}{({#2}\|{#3})}}}

\newcommand{\IdistLn}[3]    {\mm{{d}_{#1}{({#2},{#3})}}}
\newcommand{\IdistSq}[3]    {\mm{{d_{#1}^2}{({#2},{#3})}}}
\newcommand{\Idist}[1]      {\mm{{d}_{#1}}}
\newcommand{\DivF}[1]       {\mm{{D}_{#1}}}
\newcommand{\DivFInv}[1]    {\mm{{D}_{#1}^{-1}}}
\newcommand{\Ball}[3]       {\mm{{B}_{{#2}}{({#3})}}}

\newcommand{\diff}          {\mm{\rm \,d}}
\newcommand{\image}[1]      {\mm{\rm im}\,{#1}}

\newcommand{\ee}            {\mm{\varepsilon}}

\newcommand{\Edist}[2]      {\mm{\|{#1}-{#2}\|}}
\newcommand{\Maxdist}[2]    {\mm{\|{#1}-{#2}\|_\infty}}
\newcommand{\bottleneck}[2] {\mm{{W_\infty}{({#1},{#2})}}}
\newcommand{\interleave}[2] {\mm{{I}{({#1},{#2})}}}
\newcommand{\Dgm}[1]        {\mm{\rm Dgm}{({#1})}}

\newcommand{\Skip}[1]       {}

\title{Topological Data Analysis in Information Space\footnote{This research
       is partially supported by the Office of Naval Research, through grant
no. N62909-18-1-2038, and the DFG Collaborative Research Center TRR 109,
       `Discretization in Geometry and Dynamics', through grant no.\ I02979-N35 of the
       Austrian Science Fund (FWF).}}
\titlerunning{Topological Data Analysis in Information Space} 

\author{Herbert Edelsbrunner}{IST Austria (Institute of Science and Technology Austria)
  \\Am Campus 1, 3400 Klosterneuburg, Austria}{edels@ist.ac.at}{}{}{}
\author{\v{Z}iga Virk}{Faculty of Computer and Information Science, University of Ljubljana
  \\Vecna pot 113, 1000 Ljubljana, Slovenia}{ziga.virk@fmf.uni-lj.si}{}{}{}
\author{Hubert Wagner}{IST Austria (Institute of Science and Technology Austria)
  \\Am Campus 1, 3400 Klosterneuburg, Austria}{hwagner@ist.ac.at}{}{}{}

\authorrunning{H. Edelsbrunner and \v{Z}. Virk and H. Wagner} 

\Copyright{H. Edelsbrunner and \v{Z}. Virk and H. Wagner}



\begin{document}
\maketitle

\begin{abstract}
Various kinds of data are routinely represented as discrete probability distributions. Examples include text documents summarized by histograms of word occurrences and images represented as histograms of oriented gradients. Viewing a discrete probability distribution as a point in the standard simplex of the appropriate dimension, we can understand collections of such objects in geometric and topological terms. Importantly, instead of using the standard Euclidean distance, we look into dissimilarity measures with information-theoretic justification, and we develop the theory needed for applying topological data analysis in this setting. In doing so, we emphasize constructions that enable usage of existing computational topology software in this context.
\end{abstract}


\section{Introduction}
\label{sec1}

The field of \emph{computational topology} has a very short history,
at least if compared with standard subfields of mathematics \cite{EdHa10}.
Starting as a broadening of the geometric tool-set to answer low-dimensional
topology questions, it soon extended its scope to include
topological problems in high-dimensional data analysis,
giving rise to the subfield of \emph{topological data analysis} (TDA)~\cite{Car09}.
At the very foundation of topology is the notion of neighborhood, which
in practice is often
constructed from the Euclidean metric in real space.
One reason for its popularity is the extensive tool-set
associated with this metric, but there is evidence that this choice
is often suboptimal~\cite{Hua08}.
Other popular options are the Hamming distance, preferred for its
simplicity, and discrete metrics, because they minimize the structural
prerequisites.

While the choice of distance is important, there is a paucity in
our understanding of the consequences.
We contribute to the study by connecting distances with
information-theoretic foundations \cite{CsKo11}
to the tool-set of topological data analysis.
These distances include
the \emph{Fisher information metric} \cite{Aki79},
and the \emph{relative entropy},
also known as \emph{Kullback--Leibler divergence} \cite{KuLe51}.
The former is obtained by integrating
the square root of the relative entropy along shortest paths.
Restricting the points to the standard simplex,
we get a metric between discrete probability
distributions, which has information-theoretic significance.
While the latter is not a distance in the strict sense,
it plays an important role in many application areas as preferred measurement 
between probability distributions. Kullback--Leibler divergence is also an example of the broader class of Bregman divergences \cite{Bre67},
which have been studied from a geometric point of view
in \cite{BNN10,NiNo06}.
To clearly distinguish the spaces equipped with dissimilarity measures based on
information-theoretic notions from the standard Euclidean setting, 
we call them \emph{information spaces}.

Our interest in information-theoretic distances complements the 
common theme of incorporating statistical tools into the 
topological data analysis pipeline.
Like-minded efforts advocate
that results obtained with topological methods need calibration
through statistics, and that new developments in both fields
will be needed to achieve this \cite{Adl14}.
The goal of this paper is more modest:
the import of concepts in information theory
into the inner workings of the topological tools.
This can be compared with the \emph{persistent entropy}
introduced by Rucco and collaborators \cite{MRST15}
to sharpen the sensitivity of persistent homology in the
assessment of network organization.
Unrelated to this thought but important to our effort
is the work of Antonelli and collaborators \cite{Antonelli} on the isometry between
information space and Euclidean space,
which we recast in Section \ref{sec3}.
We follow up with a list of concrete contributions the reader finds in this paper:
\begin{itemize}\denselist
  \item In Section \ref{sec2}, we prove that relative entropy balls are convex;        
    in contrast, the balls based on the Burg entropy,
    which is a popular in speech recognition, are convex only for small radii.
  \item In Section \ref{sec3}, we give an explicit description of
    the Fisher information metric
    derived from the relative entropy and of Antonelli's isometry,
    prove that Fisher information balls are convex, and generalize
    the Antonelli isometry to all decomposable Bregman divergences. 
  \item  In Section \ref{sec4}, we prove tight bounds relating the Jensen--Shannon divergence
    with the Fisher information metric,
    recalling that the square root of the former is a metric \cite{EnSc03}.
\end{itemize}
In a final technical section, we tie things together by
explaining how these notions of distance are used to compute the persistence diagrams
of data in this space.
We compare the results for the three notions of distance studied in this paper:
the \emph{relative entropy}, the \emph{Fisher information metric},
and the \emph{Jensen--Shannon divergence}. 
Due to space restrictions, considerations for the Burg entropy are moved to Appendices.

\paragraph{Scope of the paper.}
Apart from reporting the above technical results, our aim is to provide a self-contained reference for researchers interested in the theoretical and applied side of information theory in the context of computational topology. To this end, we provide 
intuitive explanations of basic topological objects measured with information-theoretic distances.

\paragraph{Prospects in data analysis.}
We briefly position our work with respect to modern data analysis.
In general, topological data analysis promises tools for understanding and comparing global properties of data, also in the challenging high dimensional case. We see growing need for such tools, as deep neural networks, dimensionality 
reduction methods, and generative adversarial networks are becoming commonplace in many application areas. Topological data analysis offers supervision and validation, promising to answer questions such as the following: How is the connectivity affected by the dimensionality reduction? What are the (topological) obstructions for obtaining a high quality reduction for a target space of a given dimension? How does the space of data generated by generative adversarial networks compare to the space of 'natural' data? Finally, can we provide a topological description of the workings of the modern neural networks? 

Questions about local properties are of interest as well, such as about the dimensionality of the data. This is a topological question, but was only studied in Euclidean spaces~\cite{BeWaMu12}. How does 
changing the distance measure affect the results, especially if the change is to a non-metric measure of dissimilarity? 
Of course, the above questions are ambitious. In this paper we focus on providing basic tools, allowing 
us to start thinking about them, free from the Euclidean, or more generally the metric assumptions.

A staple example of topological data analysis is the experimental study of natural images,
as viewed through the lens of topology of high contrast patches~\cite{Car09}. 
This is an instance of a standard technique in image recognition, namely embedding images in an appropriate feature space that captures the 
relevant properties but reduces data dimensionality.
In particular, an image can be associated with 
a histogram that describes its characteristic properties. 
The simplest example maps each image to the histogram of pixel intensity values~\cite{color}. A more elaborate technique
is the \emph{histogram of oriented gradients} approach~\cite{dalal2005histograms}. 
Going beyond images, we find that text documents are expressed by histograms of word occurrences, or term-vectors~\cite{Hua08}; and that  time-series in computational neuroscience are characterized by the peristimulus time histograms~\cite{neuro}. Having mapped a collection of objects to a feature space, we use geometric tools to quickly compare pairs of objects, efficiently search for similar objects given a query object, etc.

We remark that in these cases the data naturally lives 
in the space of discrete probability distributions --- or the information space --- where the direct usage of the Euclidean metric is often not effective. Instead, methods with information-theoretic justification are likely more suitable. Such methods, and their interplay with topology, are the topic of this paper.

\paragraph{Overview.}
To put our results in perspective,
we mention that this paper is third in a series.
The entry point is~\cite{EdWa16}, where Bregman divergences are shown to be
compatible with basic topological tools.
In subsequent work~\cite{EdViWa17}, we 
studied properties of Bregman balls, in particular bounding the location of first intersection
of such balls. Our previous work was concerned with shared properties of Bregman divergences; in this
work we focus on information-theoretic distances based on Shannon's entropy. 

While working directly with the Kullback--Leibler divergence is ideal from an information-theoretic standpoint, 
its lack of certain properties, such as the triangle inequality, prohibits its use with existing computational geometry and topology tools. This prompts the study of metrics derived from Bregman divergences in general,
and Kullback--Leibler divergence in particular. For the latter these metrics include the Fisher metric and Jensen-Shannon metric. 

Studying properties of these metrics is the main focus of this paper, which is structured as follows: 
Section \ref{sec2} presents the background on Shannon entropy
and relative entropy.
Section \ref{sec3} relates the Fisher information metric to the relative
entropy and to the Euclidean metric.
Section \ref{sec4} proves inequalities relating the Jensen--Shannon divergence
with the Fisher information metric.
Section \ref{sec5} uses persistent homology to compare
these notions of distance if applied to data.
Section \ref{sec6} concludes this paper.
In addition, Appendices~\ref{app:burgdiv},~\ref{app:burginf},~\ref{app:br}
consider distance measures based on the Burg entropy.



\section{Entropy and Relative Entropy}
\label{sec2}

Starting with the Shannon entropy, we define the relative entropy of an ordered
pair of points, and provide an intuitive explanation of this measurement.
Considering the ball consisting of all points for which
the relative entropy from the center does not exceed a given threshold,
we prove that it is convex, in all dimensions.
This is in contrast to the balls defined by the Burg entropy,
which we show are nonconvex unless they are small (see Appendix~\ref{app:burgdiv}).

\begin{figure}[bht]
  \centering \resizebox{!}{1.4in}{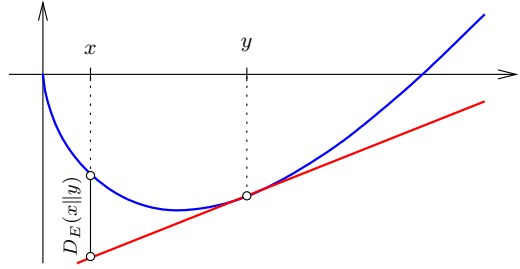}
  \caption{The graph of the Shannon entropy, the graph of its best linear
    approximation at $y$, and the relative entropy from $x$ to $y$.}
  \label{fig:relent}
\end{figure}

\paragraph{Convex function and divergence.}
We write $\Rspace_+$ for the set of positive real numbers,
$\Rspace_+^n$ for the positive orthant in $n$ dimensions,
and $x = (x_1, x_2, \ldots, x_n) \in \Rspace_+^n$ for a typical
point in the orthant.
Its components satisfy $x_i > 0$ for $1 \leq i \leq n$.
The (negative) \emph{Shannon entropy} is the
function $E \colon \Rspace_+^n \to \Rspace$ defined by mapping $x$ to 
\begin{align}
  E(x)             &=  \sum\nolimits_{i=1}^n [ x_i \ln x_i - x_i ] .
    \label{eqn:Shannon}
\end{align}
It is strictly convex and infinitely often differentiable;
see Figure \ref{fig:relent} for the graph of the function in one dimension.
The \emph{relative entropy} from $x$ to $y$
is the difference between $E$ and the best linear approximation of $E$ at $y$,
both evaluated at $x$:
\begin{align}
  \Bdist{E}{x}{y}  &=  E(x) - [E(y) + \scalprod{\nabla E(y)}{x-y}]
    \label{eqn:divergence} \\
                   &=  \sum\nolimits_{i=1}^n
                       [x_i \ln \tfrac{x_i}{y_i} - x_i + y_i].
    \label{eqn:rel-entropy}
\end{align}
The relative entropy is also known
as the \emph{Kullback--Leibler divergence}; see \cite[page 57]{AmNa00}.
The construction of the relative entropy
is illustrated in Figure \ref{fig:relent},
making it obvious that $\Bdist{E}{x}{y}$ is neither symmetric in its two
arguments, nor does it satisfy the triangle inequality.
Notwithstanding, the primary interpretation of the relative entropy
is as a measure of distance. Intuitively, the relative entropy from
a distribution $x$ to another distribution $y$ measures the average \emph{surprisal} when we expect 
items according to $y$ but observe items according to $x$. More technically,
this surprisal is expressed as the loss of coding efficiency, 
namely the expected number of extra bits needed to encode 
the messages from $x$ using a code optimized for $y$. In practice,
it quantifies how well $y$ approximates $x$, and later we use this interpretation
to understand the meaning of relative-entropy balls and their intersections.

\noindent {\sc Remark 1:}
The definition of the relative entropy is an example
of a more general construction.
Letting $\Domain \subseteq \Rspace^n$ be open and convex,
a function $F \colon \Domain \to \Rspace$ is of \emph{Legendre type}
if it is strictly convex, differentiable, and the gradient diverges
when we approach a boundary point of $\Domain$.
Clearly, the Shannon entropy is of Legendre type.
Named after Lev Bregman \cite{Bre67}, the corresponding
\emph{Bregman divergence} is defined by substituting $F$ for $E$ in
\eqref{eqn:divergence}.

\medskip \noindent {\sc Remark 2:}
If we restrict a Legendre type function to an affine subspace,
we get another Legendre type function and therefore another divergence.
An important example is the \emph{standard $(n-1)$-simplex},
$\Delta = \Delta^{n-1}$,
which is the intersection of $\Rspace_+^n$ with the $(n-1)$-plane
defined by $\sum_{i=1}^n x_i = 1$.
Since $\Rspace_+^n$ is open so is $\Delta$.
Every point of $\Delta$ can be interpreted as a discrete probability
distribution for $n$ elements.
The Shannon entropy and the relative entropy are defined as in
\eqref{eqn:Shannon} and \eqref{eqn:rel-entropy},
except for a smaller domain.

\paragraph{Decomposability and convexity.}
We note that a function $F \colon \Domain \to \Rspace$
whose restrictions to fixed coordinate values
are convex is not necessarily convex.
An example is the function $F(x) = \prod_{i=1}^n x_i$.
On the other hand, if $F$ is \emph{decomposable}:
$F(x) = \sum_{i=1}^n F_i (x_i)$,
then the convexity of the components implies the convexity of $F$.
\begin{lemma}[Composable Convexity]
  \label{lem:ComposableConvexity}
  Let $\Domain \subseteq \Rspace^n$ be convex and
  $F \colon \Domain \to \Rspace$ a decomposable function.
  The components of $F$ are convex if and only if $F$ is convex.
\end{lemma}
\myproof
  It is clear that the convexity of $F$ implies the convexity of
  its components.
  We thus limit ourselves to proving the other implication.
  Let $x, y \in \Domain$, let $0 \leq \lambda \leq 1$,
  and let $u = (1-\lambda) x + \lambda y$
  be the corresponding convex combination.
  Since the components of $F$ are convex, by assumption, we have
  $F_i (u_i) \leq (1-\lambda) F_i (x_i) + \lambda F_i (y_i)$,
  for every $1 \leq i \leq n$.
  Taking sums on the left and on the right, we get
  $F (u)  \leq  (1-\lambda) F (x) + \lambda F (y) ;$
  that is: $F$ is convex as claimed.
\myeop

\noindent {\sc Remark 3:}
While nonconvex components imply a nonconvex function, they do not imply
nonconvex sublevel sets.
This stronger implication is valid for \emph{strongly decomposable}
functions $F$, by which we mean that all components are the same:
$F_i = F_j$ for all $1 \leq i, j \leq n$.
To see this, let the component $f$ of $F$ be nonconvex,
assume $f$ is nonnegative, and let $a < b$ such that
$f (\tfrac{a+b}{2}) > \tfrac{1}{2} f (a) + \tfrac{1}{2} f (b)$.
Supposing $1$ is in the domain of $f$, we set
$x = (a, b, 1, \ldots, 1)$ and $y = (b, a, 1, \ldots, 1)$ and note that
$\tfrac{x+y}{2} = (\tfrac{a+b}{2}, \tfrac{a+b}{2}, 1, \ldots, 1)$.
Hence,
$F(\tfrac{x+y}{2})  =  2 f (\tfrac{a+b}{2}) + (n-2) f (1)   
                    >  f(a) + f(b) + (n-2) f(1)$,
which is $\tfrac{1}{2} [F(x) + F(y)]$.
Setting $r^2 = F(x) = F(y)$, we see that $F^{-1} [0,r^2]$
is a nonconvex subset of $\Rspace^n$.

\paragraph{Convexity of entropy balls.}
To study the local behavior of a divergence, we fix a point $x \in \Domain$
and call $\DivF{x} \colon \Domain \to \Rspace$ defined by
$\DivF{x} (y) = \Bdist{F}{x}{y}$ the \emph{divergence function} of $x$.
Its sublevel sets are balls of points whose divergence from $x$ is
bounded by a non-negative threshold:
\begin{align}
  \Ball{F}{r}{x}  &=  \DivFInv{x} [0,r^2]
                   =  \{ y \in \Domain \mid \Bdist{F}{x}{y} \leq r^2 \} .
\end{align}
Assuming $F$ is the Shannon entropy, we call
$\Ball{F}{r}{x}$ an \emph{entropy ball}. Observe that 
such a ball is the set of all probability distributions 
that approximate the center distribution, $x$, with a loss of at most $r^2$ bits. 
We recall that in practice each distribution characterizes an object, perhaps an image or a text document.
In this case, it is clear that a ball represents a collection of similar objects, for a chosen similarity threshold $r^2$.
Furthermore, the point of first common intersection of a collection of balls
of radius $r^2$ is the best approximation for a set of objects represented by the 
centers of the balls; each center can be approximated with a loss 
of at most $r^2$ bits.

For general Legendre type functions, the balls are not necessarily convex,
and we will encounter nonconvex balls shortly.
However, entropy balls are necessarily convex.
\begin{theorem}[Convex Entropy Balls]
  \label{thm:ConvexEntropyBalls}
  Let $E \colon \Rspace_+^n \to \Rspace$ be the Shannon entropy.
  Then the entropy ball $\Ball{E}{r}{x}$ is convex for every
  $x \in \Rspace_+^n$ and every $r^2 \geq 0$.
\end{theorem}
\myproof
  Since the Shannon entropy is strongly decomposable,
  so is the divergence function of every point $x \in \Rspace_+^n$:
  $\DivF{x} (y)  =  E(x) - E(y) - \scalprod{\nabla E(y)}{x-y}      
      =  \sum\nolimits_{i=1}^n [ x_i \ln x_i - x_i \ln y_i - x_i + y_i ]$,
  in which the $y_i$ are variables and the $x_i$ are constants.
  Each component is convex because $- \ln t$ is convex.
  By Lemma \ref{lem:ComposableConvexity}, this implies that
  $\DivF{x}$ is convex and by Remark 3
  that it has only convex sublevel sets.
\myeop

\section{Fisher Information Metric}
\label{sec3}

Like any other twice differentiable Legendre type function,
the Shannon entropy
induces a metric that integrates infinitesimal steps
along shortest paths.
This metric has an isometry to Euclidean space,
which facilitates topological methods applied to data in information space.
Analogous study based on the Burg entropy can be found in Appendix~\ref{app:burginf}.

\paragraph{From divergence to distance.}
Recall that the Shannon entropy is twice differentiable, so the
\emph{Hessian matrix} of second derivatives at any point $y \in \Rspace_+^n$
is well defined:
\begin{align}
  H_E(y)  &=  \left[ \frac{\partial^2 E}{\partial x_i \partial x_j} (y)
            \right]_{1 \leq i, j \leq n} .
\end{align}
This symmetric square matrix defines
a scalar product at $y$ by mapping two vectors $u, v \in \Rspace^n$ to
$\scalprod{u}{v}_y = \tfrac{1}{2} \, u^T H_E(y) v$.
The corresponding Riemannian metric is called the
\emph{Fisher information metric}:
$\Idist{E} \colon \Rspace_+^n \times \Rspace_+^n \to \Rspace$.
Given points $x, y \in \Rspace_+^n$,
the distance is the length of the shortest path
$\gamma \colon [0,1] \to \Rspace_+^n$ with $\gamma(0) = x$
and $\gamma(1) = y$, in which the length is
\begin{align}
  \Length{\gamma}  &=  \int_{t=0}^1
    \sqrt{ \scalprod{\dot\gamma (t)}{\dot\gamma (t)}_{\gamma(t)} } \diff t
                   =  \int_{t=0}^1
    \sqrt{ \tfrac{1}{2} \, \dot\gamma(t)^T H_E(\gamma(t)) \dot\gamma(t)} \diff t .
\end{align}
Since the Shannon entropy is decomposable, the only non-zero entries
in the Hessian matrix are the second derivatives of
the component functions in the diagonal:
\begin{align}
  H_E(y)  &=  \left[ \begin{array}{cccc}
              1/y_1   &  0      &  \ldots  &  0            \\
              0       &  1/y_2  &  \ldots  &  0            \\
              \vdots  &  \vdots &  \ddots  &  \vdots       \\
              0       &  0      &  \ldots  &  1/y_n
            \end{array} \right] .
  \label{eqn:Hessian}
\end{align}
Accordingly, the formula for the length of a smooth path simplifies to
\begin{align}
  \Length{\gamma}  &=  \int_{t=0}^1
    \sqrt{ \tfrac{1}{2} \sum\nolimits_{i=0}^n
           \tfrac{\dot\gamma_i (t)^2}{\gamma_i (t)} } \diff t ,
\end{align}
in which $\gamma_i (t)$ and $\dot\gamma_i (t)$ are the $i$-th components
of the curve and its velocity vector at $t$.

\medskip \noindent {\sc Remark 4:}
The construction of the metric can be restricted to any smooth submanifold
of the domain.
Since this limits the set of available paths,
the distance between two points cannot be less than without the restriction.
For example, we may restrict the Fisher information metric in $\Rspace_+^n$
to the standard simplex:
$\Idist{E|_\Delta} \colon \Delta \times \Delta \to \Rspace$.
While the relative entropy in $\Delta$ is the restriction
of the relative entropy in $\Rspace_+^n$, this is not the case
for the Fisher information metric.

\paragraph{Euclidean isometries.}
As discovered by Antonelli \cite{Antonelli}, there is a simple
diffeomorphism that maps $\Rspace_+^n$ equipped with the Fisher information metric
to $\Rspace_+^n$ equipped with the Euclidean metric.
Such an isometry exists for every length metric defined as explained
by a twice differentiable decomposable Legendre type function.
We explain its construction in this more general setting.
Suppose $\varphi \colon \Domain \to \Domain_2$ is a diffeomorphism.
Given a metric $\Idist{2} \colon \Domain_2 \times \Domain_2 \to \Rspace$,
we get an \emph{induced metric},
$\Idist{\varphi} \colon \Domain \times \Domain \to \Rspace$,
defined by $\IdistLn{\varphi}{x}{y}
          = \IdistLn{2}{\varphi(x)}{\varphi(y)}$.
By construction, $\varphi$ is an isometry between the two metric spaces.
We are interested in the case in which $\Idist{2}$ is the
Euclidean metric.
\begin{lemma}[Euclidean Isometry]
  \label{lem:EuclideanIsometry}
  Let $\Domain$ be an open convex subset of $\Rspace^n$ and
  $F \colon \Domain \to \Rspace$ a twice differentiable
  decomposable Legendre type function.
  Then there exists a decomposable isometry
  $\varphi \colon \Domain \to \Domain_2 \subseteq \Rspace^n$
  connecting the length metric defined by $F$ in $\Domain$
  with the Euclidean metric in $\Domain_2$.
\end{lemma}
\myproof
  We construct a decomposable diffeomorphism
  $\varphi \colon \Domain \to \Domain_2$
  such that the length metric $\Idist{F}$ is induced by $\varphi$ and
  the Euclidean metric on $\Domain_2$.
  Write $\varphi_i (x_i)$ for the components of the diffeomorphism
  and $\sum_{i=1}^n (\! \diff u_i)^2$ for the differential form of the
  Euclidean metric in $\Domain_2$.
  Setting $u_i = \varphi_i (x_i)$
  so that $\diff u_i = \varphi_i' (x_i) \diff x_i$, we get
  \begin{align}
    \sum\nolimits_{i=1}^n (\! \diff u_i )^2
      &=  \sum\nolimits_{i=1}^n \varphi_i' (x_i)^2 (\! \diff x_i)^2
       =  \sum\nolimits_{i=1}^n \tfrac{1}{2} F_i'' (x_i) (\! \diff x_i)^2 ,
  \end{align}
  in which the last sum is the differential form of the metric
  defined by $F$, and the second equality is required to get an isometry.
  This condition simplifies to
  \begin{align}
    \varphi_i' (x_i) &= \sqrt{ \tfrac{1}{2} F_i'' (x_i) } ,
      \label{eqn:diffeoprime}
  \end{align}
  for $1 \leq i \leq n$.
  Since $F$ is strictly convex, we have $F_i''(x_i) > 0$ for every
  component function, so this equation has a solution.
\myeop

\begin{figure}[hbt]
  \centering \vspace{0.1in}
    \includegraphics[width=0.33\textwidth]{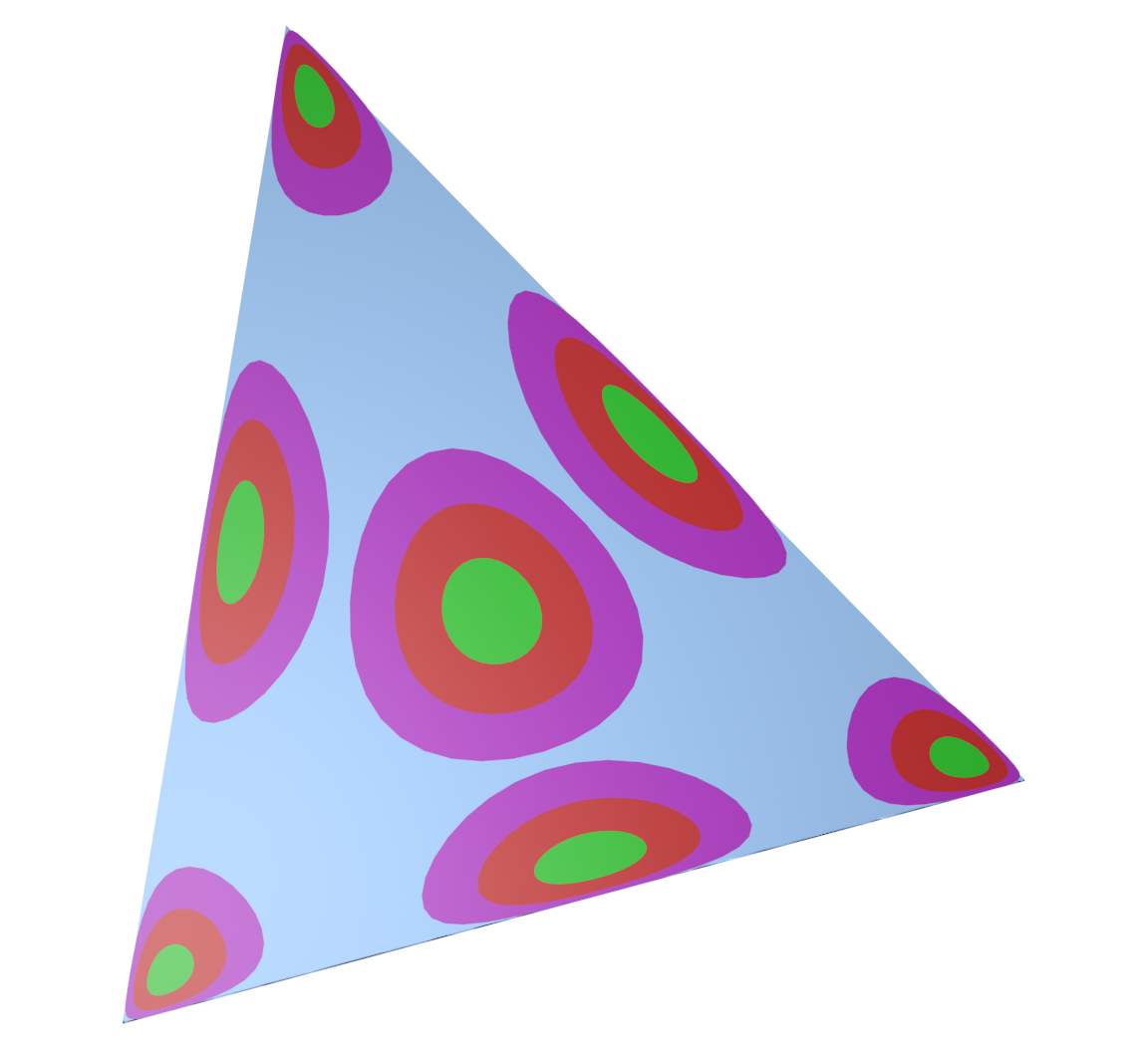}
    \hspace{0.00in}
    \includegraphics[width=0.33\textwidth]{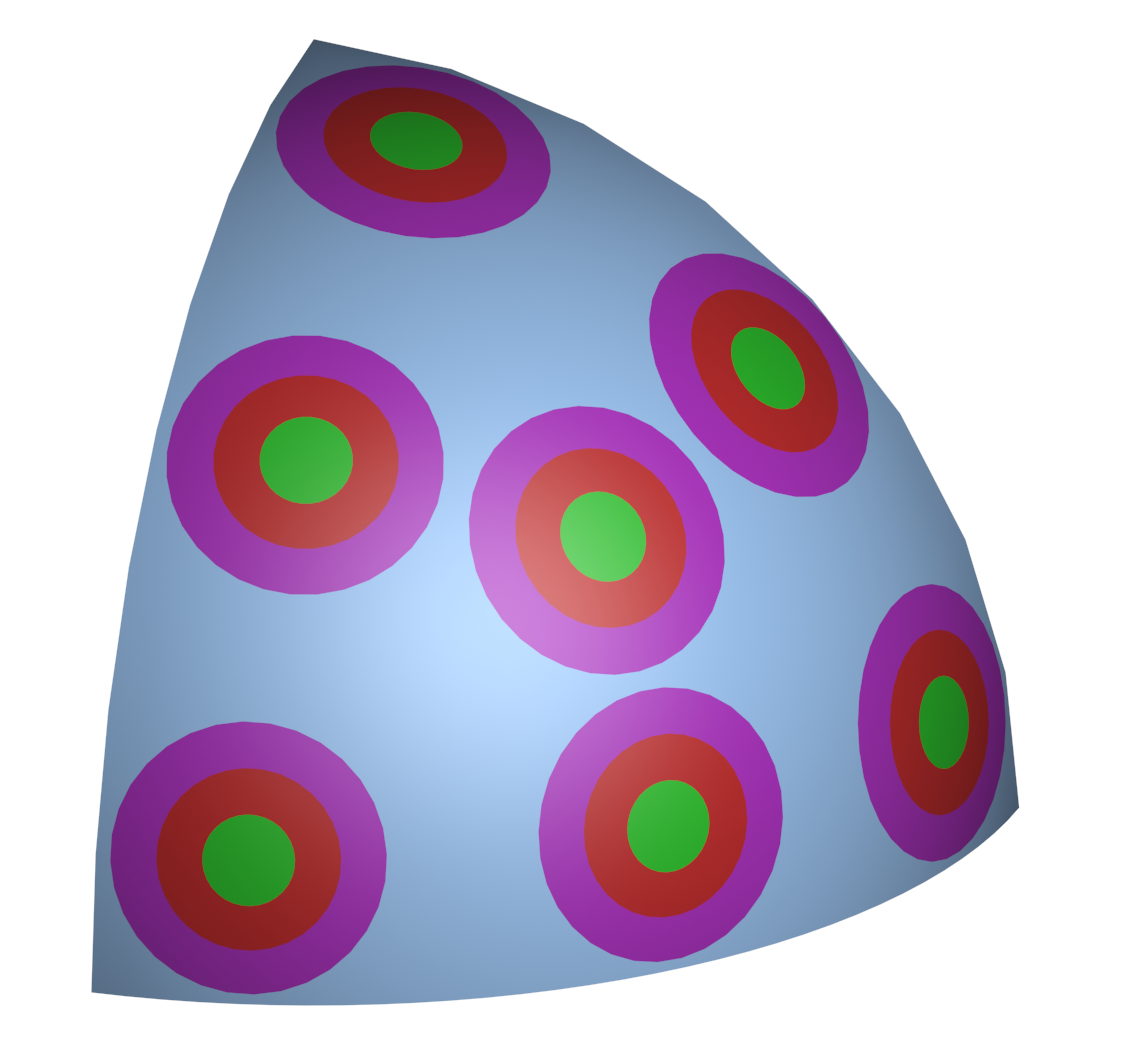}
    \vspace{0.05in}
  \caption{\emph{Left:} three Fisher information disks each for seven points
    inside the standard triangle.
    \emph{Right:} the images of the disks in the positive
    portion of the sphere.
    For aesthetic reasons, we scale the sphere
    by a factor $1 / \sqrt{2}$ relative to the triangle.}
  \label{fig:infspace}
\end{figure}

\paragraph{From Fisher information to Euclidean distance.}
We make use of Lemma \ref{lem:EuclideanIsometry} by constructing the isometry
that relates the Fisher information metric in $\Rspace_+^n$ and in
$\Delta = \Delta^{n-1}$ with the Euclidean metric in
$\Rspace_+^n$ and in $\sqrt{2} \Sspace_+^{n-1}$,
the latter being our notation for the positive orthant of the $(n-1)$-sphere
with radius $\sqrt{2}$ centered at the origin in $\Rspace^n$.
As mentioned before, this isometry is well known \cite{Aki79,Antonelli},
and we give the construction for completeness.
\begin{theorem}[Fisher Information to Euclidean Distance]
  \label{thm:FItoED}
  Let $E \colon \Rspace_+^n \to \Rspace$ be the Shannon entropy,
  and $x, y \in \Rspace_+^n$.
  Then
  \begin{align}
    \IdistLn{E}{x}{y}  &=  \sqrt{ 2 \sum\nolimits_{i=1}^n
          \left( \sqrt{x_i} - \sqrt{y_i} \right)^2 } ,
      \label{eqn:Edistance1} \\
    \IdistLn{E|_{\Delta}}{x}{y}
        &=  \sqrt{2} \arccos \sum\nolimits_{i=1}^n \sqrt{x_i} \sqrt{y_i} ,
      \label{eqn:Edistance2}
  \end{align}
  where we assume $x, y \in \Delta \subseteq \Rspace_+^n$
  in \eqref{eqn:Edistance2}.
  Furthermore, the balls under $\Idist{E}$
  and $\Idist{E|_\Delta}$ are convex.
\end{theorem}
\myproof
  We prove both equations with the diffeomorphism
  $\varphi \colon \Rspace_+^n \to \Rspace_+^n$ that distorts the
  Fisher information metric to the Euclidean metric.
  Since $E$ decomposes into identical components, so does $\varphi$.
  By Equation \eqref{eqn:diffeoprime} in the proof of
  Lemma \ref{lem:EuclideanIsometry}, the derivative of this component
  satisfies $p'(t) = 1 / {\sqrt{2t}}$.
  We get $p(t) = \sqrt{2t}$ by integration.
  As illustrated in Figure \ref{fig:infspace}, the diffeomorphism
  maps a point $x$ with coordinates $x_i > 0$ to the point $\varphi (x)$
  with coordinates $p(x_i) = \sqrt{2 x_i} > 0$.
  The squared Euclidean distance between $\varphi(x)$ and $\varphi(y)$ is
  $\sum_{i=1}^n ( \sqrt{2 x_i} - \sqrt{2 y_i} )^2$,
  which implies \eqref{eqn:Edistance1}.

  The image of the standard simplex, $\varphi (\Delta)$,
  consists of all points with coordinates $u_i = p(x_i) > 0$ that satisfy
  $u_1^2 + u_2^2 + \ldots + u_n^2  =  2$.
  This is the equation of the sphere with radius $\sqrt{2}$
  centered at the origin in $\Rspace^n$.
  Denoting the positive portion of this sphere by $\sqrt{2} \Sspace_+^{n-1}$,
  we get an isometry
  $\varphi|_\Delta \colon \Delta \to \sqrt{2} \Sspace_+^{n-1}$.
  Given points $u, v \in \sqrt{2} \Sspace_+^{n-1}$,
  the shortest spherical path connecting them is a portion
  of the great-circle that passes through $u$ and $v$.
  Its length is $\sqrt{2}$ times the angle between the vectors
  $u, v \in \Rspace_+^n$.
  This implies \eqref{eqn:Edistance2}.

  To prove convexity, let $x \in \Rspace_+^n$, consider the function
  $d_x \colon \Rspace_+^n \to \Rspace$ defined by $d_x(y) = \IdistSq{E}{x}{y}$,
  and note that $d_x^{-1} [0, r^2]$ is the ball with center $x$
  and radius $r \geq 0$ under the Fisher information metric.
  It is the preimage of the function $f_u \colon \Rspace_+^n \to \Rspace$
  defined by $f_u(v) = \sum_{i=1}^n (v_i - u_i)^2$,
  with $u_i = \sqrt{2 x_i}$ and $v_i = \sqrt{2 y_i}$, for $1 \leq i \leq n$.
  Hence, $d_x$ is decomposable with
  $d_x(y) = 2 \sum_{i=1}^n (\sqrt{y_i} - \sqrt{x_i})^2$.
  The components are convex because $- \sqrt{t}$ is convex.
  By Lemma \ref{lem:ComposableConvexity}, $d_x$ is convex,
  and because $x$ can be any point in $\Rspace_+^n$,
  all balls under $\Idist{E}$ are convex.
  To finally see that every ball under $\Idist{E|_\Delta}$ is convex,
  we note that it is the preimage of the intersection between
  $\sqrt{2} \Sspace_+^{n-1}$ and a Euclidean ball.
  This preimage is the intersection between $\Delta$ and a ball
  under $\Idist{E}$, which is again convex.
\myeop

\noindent {\sc Remark 5:}
Using the isometry specified in Theorem \ref{thm:FItoED},
we can map data from Fisher information space to Euclidean space and apply
standard mathematical and computational tools there.
Furthermore, the presented derivation works for any Bregman
divergence, provided it is decomposable, which ensures the Hessian is diagonal. 
See Appendix~\ref{app:burginf} for a version based on the Itakura-Saito divergence.

\section{Information and Loss}
\label{sec4}

In this section, we reverse the direction and measure divergences
\emph{to} a point.
We use these readings to define the Jensen--Shannon divergence or \emph{loss}, 
whose square root we compare to the Fisher information metric. Furthermore, 
in Appendix~\ref{app:br}, we consider the Burbea-Rao divergence that is based
on the Burg entropy.

\paragraph{Jensen--Shannon divergence.}
The \emph{Jensen--Shannon divergence} of a
pair $x, y \in \Rspace_+^n$ is the average relative entropy from $x$ and $y$
to the average of the two points:
\begin{align}
  \JSLs{x}{y}  &=  \tfrac{1}{2} [\Bdist{E}{x}{\mu}+\Bdist{E}{y}{\mu}] 
                =  \tfrac{1}{2} [E(x) + E(y)] - E(\mu)                \\
               &=  \tfrac{1}{2} \sum\nolimits_{i=1}^n
                  [x_i \ln \tfrac{2x_i}{x_i+y_i}
                 + y_i \ln \tfrac{2y_i}{x_i+y_i}] ,
    \label{eqn:JS-3}
\end{align}
in which $\mu = \tfrac{x+y}{2}$ and we get (\ref{eqn:JS-3}) by noting
that $x + y - 2 \mu = 0$.
As pointed out in \cite{EnSc03}, it measures the loss of efficiency when
we encode signals from two probability distributions using their average.
We contrast this quantity with the radius at which two entropy balls intersect.
In the latter case, we obtain the worst-case loss of efficiency, as opposed to the average.

If we substitute any other point $z \in \Rspace_+^n$ for $\mu$,
the average relative entropy increases.
To prove this, we define
$f(z) =  \tfrac{1}{2} [\Bdist{E}{x}{z} + \Bdist{E}{y}{z}]$,
noting that $f(\mu) = \JSLs{x}{y}$.
The following lemma can also be found in \cite{BMDG05}.
\begin{lemma}[Local Minimum]
  \label{lem:LocalMinimum}
  Let $\mu$ be the average of the points $x, y \in \Rspace_+^n$.
  Then $f(\mu) \leq f(z)$ for every $z \in \Rspace_+^n$,
  with equality if and only if $z = \mu$.
\end{lemma}
\myproof
  Computing $f(z) - f(\mu)$ from the definition,
  most terms cancel and only
  $E(\mu) - E(z) - \scalprod{\nabla E(z)}{\mu-z}$ remains,
  which implies $f(z) - f(\mu) = \Bdist{E}{\mu}{z}$.
  It is non-negative by definition and $0$ if and only if $z = \mu$.
\myeop

As proved in \cite{EnSc03}, $\Rspace_+^n$ together with the
square root of the Jensen--Shannon divergence is a metric space.
It is however not a \emph{length metric space} because
it does not agree with the corresponding \emph{intrinsic metric},
in which the length of a path is measured by taking infinitesimal
steps; see also Section \ref{sec3}.
It is not difficult to see that the intrinsic metric defined by the
Jensen--Shannon divergence is half the intrinsic metric defined
by the relative entropy, which is of course the Fisher information metric.
Being a metric, the square root of the Jensen--Shannon divergence
satisfies the triangle inequality, which implies that it is bounded
from above by the corresponding intrinsic metric.

\paragraph{Jensen--Shannon divergence versus Fisher information metric.}
Since the Fisher information metric is twice the intrinsic metric
defined by the Jensen--Shannon divergence,
we get $4 \, \JSLs{x}{y} \leq \IdistSq{E}{x}{y}$.
We will confirm this inequality with an independent argument shortly,
and prove $\tfrac{4}{\ln 2} = 5.770\ldots$
as a tight upper bound on the expansion.
\begin{theorem}[Jensen--Shannon Divergence vs.\ Fisher Information]
  \label{thm:JSLvsFI}
  The Jen\-sen--Shannon divergence and the squared Fisher information metric satisfy
  \begin{align}
    4 \, \JSLs{x}{y}  &\leq  \IdistSq{E}{x}{y} 
                    \leq  \tfrac{4}{\ln 2} \, \JSLs{x}{y} ,
      \label{eqn:JSvsInf-1} \\
    4 \, \JSLs{x}{y}  &\leq  \IdistSq{E|_\Delta}{x}{y} 
                    \leq  \tfrac{\sqrt{2} \pi}{\ln 2} \, \JSLs{x}{y} ,
      \label{eqn:JSvsInf-2}
  \end{align}
  in which \eqref{eqn:JSvsInf-1} applies to $\Rspace_+^n$
  and \eqref{eqn:JSvsInf-2} to the standard $(n-1)$-simplex.
\end{theorem}
\myproof
  Restricting $\Rspace_+^n$ to $\Delta$,
  the Jensen--Shannon divergence is unaffected,
  but the Fisher information metric expands because the shortest paths
  correspond to great-circle arcs rather than line segments
  in Euclidean space.
  The expansion is larger for longer paths, and the supremum expansion
  rate is $\pi / \sqrt{8}$, which is the length of a quarter
  great-circle over the length of the straight edge connecting its
  endpoints.
  Since $\tfrac{4}{\ln 2} \tfrac{\pi}{\sqrt{8}}
           = \tfrac{\sqrt{2} \pi}{\ln 2}$,
  \eqref{eqn:JSvsInf-1} implies \eqref{eqn:JSvsInf-2}.

  Returning to $\Rspace_+^n$, we note that both the Jensen--Shannon divergence
  and the squared Fisher information metric are decomposable.
  It suffices to prove \eqref{eqn:JSvsInf-1}
  in $n = 1$ dimension, where we write $a$ for $x$ and $b$ for $y$.
  From \eqref{eqn:Edistance1} and \eqref{eqn:JS-3}, we get
  $\IdistSq{E}{a}{b}  =  2 ( \sqrt{a} - \sqrt{b} )^2$ and
  $\JSLs{a}{b}  =  \tfrac{1}{2} [ a \ln \tfrac{2a}{a+b} + b \ln \tfrac{2b}{a+b} ]$.
  Observe that $\IdistSq{E}{Ca}{Cb} = C \, \IdistSq{E}{a}{b}$
  and $\JSLs{Ca}{Cb} = C \, \JSLs{a}{b}$ for every $C > 0$.
  We use these properties to eliminate one degree of freedom by
  setting $b = t^2 a$ to get
  \begin{align}
    \tfrac{1}{a} \, \IdistSq{E}{a}{b}  &=  \IdistSq{E}{1}{t^2}
                                     =  2 (1-t)^2 ,
      \label{eqn:denominator} \\
    \tfrac{1}{a} \, \JSLs{a}{b}        &=  \JSLs{1}{t^2}  =  \tfrac{1}{2}
      \left[ \ln \tfrac{2}{1+t^2} + t^2 \ln \tfrac{2t^2}{1+t^2} \right] .
      \label{eqn:numerator}
  \end{align}
  We will see shortly that the ratio of these two functions behaves
  monotonically within $1 < t < \infty$,
  attaining its extreme values in the two limits.
  These extreme values are the constants
  in \eqref{eqn:JSvsInf-1},
  with monotonicity proving the inequalities.
  Setting $f(t) = \JSLs{1}{t^2} / \IdistSq{E}{1}{t^2}$,
  the inequalities in \eqref{eqn:JSvsInf-1} are equivalent to
  \begin{align}
    \tfrac{\ln 2}{4}  &\leq  f(t)  \leq  \tfrac{1}{4} .
      \label{eqn:newinequality}
  \end{align}  
  Plugging the right-hand sides of \eqref{eqn:denominator} and
  \eqref{eqn:numerator} into the definition of the ratio, we get
  \begin{align}
    f(t)  &=  \tfrac{1}{4 (t-1)^2}
        \left[ \ln \tfrac{2}{t^2+1} + t^2 \ln \tfrac{2t^2}{t^2+1} \right] .
      \label{eqn:ratio}
  \end{align}
  When $t$ goes to infinity, the first term in \eqref{eqn:ratio}
  goes to $0$, while the second term goes to $\tfrac{\ln 2}{4}$,
  the lower bound in \eqref{eqn:newinequality}.
  To take the other limit, when $t$ goes to $1$,
  we use the l'Hopital rule and differentiate the numerator and
  the denominator twice.
  For the numerator, we get
  \begin{align}
    g(t)    &=  t^2 \ln t^2 - (t^2+1) \ln \tfrac{t^2+1}{2} , 
      \label{eqn:numerator1} \\
    g'(t)   &=  2t \ln t^2 - 2t \ln \tfrac{t^2+1}{2} ,
      \label{eqn:numerator2} \\
    g''(t)  &=  2 \ln t^2 - 2 \ln \tfrac{t^2+1}{2} + \tfrac{4}{t^2+1} .
      \label{eqn:numerator3}
  \end{align}
  Setting $t = 1$, we get $g(1) = g'(1) = 0$ and $g''(1) = 2$.
  Similarly, the denominator and its first derivate vanish at $t=1$,
  its second derivative is $8$,
  and $f(t)$ goes to $\tfrac{1}{4}$,
  the upper bound in \eqref{eqn:newinequality}.
  It remains to show that the ratio is monotonically decreasing,
  from $\tfrac{1}{4} = 0.25$ at $t = 1$
  to $\tfrac{\ln 2}{4} = 0.173\ldots$ at $t = \infty$.
  We accomplish this by computing the derivative of $f$ as written
  in \eqref{eqn:numerator1}:
  \begin{align}
    f'(t)  &=  \tfrac{(t \ln t^2 + t) (t-1)^2}{2 (t-1)^4} 
             - \tfrac{t^2 (t-1) \ln t^2}{2 (t-1)^4} 
             - \tfrac{(t \ln \tfrac{t^2+1}{2} + t) (t-1)^2}{2 (t-1)^4}
             + \tfrac{(t^2+1) \ln \tfrac{t^2+1}{2} (t-1)}{2 (t-1)^4} \\
           &=  \tfrac{t(t-1) \ln \tfrac{2t^2}{t^2+1}}{2 (t-1)^3}
             - \tfrac{t^2 \ln \tfrac{2t^2}{t^2+1}}{2 (t-1)^3}
             + \tfrac{\ln \tfrac{t^2+1}{2}}{2 (t-1)^3}               
            =  \tfrac{1}{2 (t-1)^3} \left[ \ln \tfrac{t^2+1}{2}
                                       - t \ln \tfrac{2t^2}{t^2+1} \right] .
  \end{align}
  To prove that $f'(t)$ is negative for all $t > 1$,
  we rewrite the numerator and compute its first two derivatives:
  $u(t)   = (t+1) \ln \tfrac{t^2+1}{2} - t \ln t^2$,
  $u'(t)  = \ln \tfrac{t^2+1}{2} + \tfrac{2t(t+1)}{t^2+1} - \ln t^2 - 2$,
  $u''(t) = \tfrac{1}{t (t^2+1)^2} [ - 2t^3 + 2t^2 + 2t - 2 ]$.
  The numerator of the second derivative factors into
  $(t-1) (- 2t^2 + 2)$, which is clearly negative for all $t > 1$.
  Note that $u(1) = u'(1) = 0$.
  Because $u''$ is negative, we have $u'(t) < 0$ and $u(t) < 0$
  for all $t > 1$.
  The latter inequality is equivalent to $f'(t) < 0$ for all $t > 1$,
  which implies that $f$ is monotonically decreasing.
  This implies \eqref{eqn:newinequality} and completes the proof.
\myeop


\section{TDA in Information Space}
\label{sec5}

While the preceding sections shed light on distances
with information-theoretic foundations,
we will now connect them to topological data analysis.
\Skip{In particular, we will construct filtrations of
  Vietoris--Rips complexes for data measured with
  the Jensen--Shannon divergence and the Fisher information metric,
  and compare the resulting persistence diagrams.
  We will contrast these constructions with filtrations
  of \v{C}ech complexes for data measured with Fisher information metric.}
We begin with the introduction of standard results from topological
data analysis, which we will then use to shed light on relation
between different notions of distance.

\paragraph{Diameter functions.}
It will be convenient to consolidate notation by writing $\Xspace$
for a topological space and $\dist \colon \Xspace \times \Xspace \to \Rspace$
for a distance.
Here we consider $\Xspace$ to either be $\Rspace_+^n$ or $\Delta^{n-1}$,
and $\dist$ to either be the square root of the Jensen--Shannon divergence,
$\sqrt{\JS}$, or the Fisher information metric, $\Idist{E}$.
When we consider the set of all subsets, we will always exclude the
empty set and therefore define
$\Power{\Xspace} = \{Q \subseteq \Xspace \mid Q \neq \emptyset\}$.
Given $\Xspace$ and $\dist$, the \emph{half-diameter function}
$\DFunction{\dist} \colon \Power{\Xspace} \to \Rspace$
is defined by mapping every non-empty subset $Q \subseteq \Xspace$ to
\begin{align}
  \DFunction{\dist} (Q) = \tfrac{1}{2} \sup_{x,y \in Q} \distFn{x}{y} ,
    \label{eqn:halfdiameter-function}
\end{align}
which we call the \emph{half-diameter} of $Q$.
Since $\dist$ is a metric, we can
define the \emph{Hausdorff distance} between two subsets of $\Xspace$
as the larger of the two directed such distances:
$\Hausdorff{\dist}{P}{Q} = \max \{ \dHausdorff{\dist}{P}{Q},
                                   \dHausdorff{\dist}{Q}{P} \}$,
in which
$\dHausdorff{\dist}{P}{Q}  =  \max_{x \in P} \min_{y \in Q} \dist (x,y)$.
This is a metric on the compact sets in $\Power{\Xspace}$,
and under it, the half-diameter function is continuous.
Finally, we note that the half-diameter function is \emph{monotonic},
that is: $\DFunction{\dist} (P) \leq \DFunction{\dist} (Q)$
whenever $\emptyset \neq P \subseteq Q \subseteq \Xspace$.

Data analysis starts with a finite set of points, $X \subseteq \Xspace$.
Let $K = \Power{X}$, noting that this is a full simplex,
and write $f \colon K \to \Rspace$ for the restriction of the
half-diameter function.
For each $r \geq 0$, we write $K_r = f^{-1} [0,r]$ for the corresponding
sublevel set, which consists of all simplices in $K$
with half-diameter $r$ or less.
By the monotonicity of $f$,
each sublevel set is a subcomplex of $K$.
We refer to $K_r$ as the \emph{Vietoris--Rips complex}
of $X$ and $\dist$ for radius $r$.
By construction, $K_r$ is a \emph{clique complex}:
a simplex belongs to $K_r$ if and only if all edges of the simplex belong to $K_r$.
It follows that the edges in $K_r$ determine the entire complex.
This has computational advantages because triangles and
higher-dimensional simplices can be treated implicitly,
computing their properties from the edges only when needed;
see e.g.\ the Ripser software of Bauer \cite{Ripser16}.
It has modeling disadvantages because triangles and higher-dimensional
simplices carry no new information.
We will return to this point when we consider \v{C}ech complexes
as an alternative construction.

\paragraph{Persistent homology.}
Persistent homology groups have been introduced in \cite{ELZ02}
to measure holes in proteins.
The concept has since found many applications
inside and outside mathematics.
The idea is however older, and the earliest reference we know is a paper
by Morse \cite{Mor40}, which develops the concept as a tool
in the study of minimal surfaces.

The concept starts with the classic notion of homology groups,
which formalize our notion of how a complex or really any
topological space is connected; see e.g.\ \cite{Hat02}.
It does so by counting the holes as ranks of abelian groups.
Choosing the coefficients in the construction of these groups
as elements of a field,
the groups are vector spaces and their ranks are the
familiar dimensions from linear algebra.
Returning to $f \colon \Power{X} \to \Rspace$,
we index the finitely many sublevel sets
consecutively from $1$ to $m$.
Mapping each complex $K_\ell$ to its $p$-th homology group,
for some fixed integer $p$,
we get a sequence of vector spaces, $\Hgroup_\ell = H_p (K_\ell)$:
$$
  \begin{array}{ccc ccc ccc ccc c}
    \ldots & \!\!\!\subseteq\!\!\! & K_{i-1} & \!\!\!\subseteq\!\!\! & K_i
           & \!\!\!\subseteq\!\!\! & \ldots  & \!\!\!\subseteq\!\!\! & K_{j-1}
           & \!\!\!\subseteq\!\!\! & K_j     & \!\!\!\subseteq\!\!\! & \ldots \\
           & \!\!\!\!\!\! & \big\downarrow   & \!\!\!\!\!\! & \big\downarrow
           & \!\!\!\!\!\! & & \!\!\!\!\!\!   & \big\downarrow
           & \!\!\!\!\!\! & \big\downarrow   & \!\!\!\!\!\! &                 \\
    \ldots & \!\!\!\to\!\!\! & \Hgroup_{i-1} & \!\!\!\to\!\!\! & \Hgroup_i
           & \!\!\!\to\!\!\! & \ldots        & \!\!\!\to\!\!\! & \Hgroup_{j-1}
           & \!\!\!\to\!\!\! & \Hgroup_j     & \!\!\!\to\!\!\! & \ldots
  \end{array}
$$
The inclusions $K_i \subseteq K_j$ induce linear maps
$h_{i,j} \colon \Hgroup_i \to \Hgroup_j$,
and by the functoriality of homology, the inclusions commute with these maps.
We call the row of complexes a \emph{filtration} and the
row of vector spaces together with the maps connecting them
a \emph{persistence module}.
This is the essential concept that permits us to measure
how long holes persist in the filtration.
To be specific, let $1 \leq i \leq j \leq m$ and call
the image of the map from $\Hgroup_i$ to $\Hgroup_j$ a
\emph{persistent homology group}, $\image{h_{i,j}} \subseteq \Hgroup_j$.
Its rank is the number of holes in $K_i$ that are still holes in $K_j$.
A homology class $\alpha$ in $\Hgroup_i$ is \emph{born} at $K_i$
if $\alpha$ does not belong to the image of the preceding group,
and it \emph{dies entering} $K_j$ if $\alpha$ enters the preimage
of that group when we go from $K_{j-1}$ to $K_j$.
More formally:
$\alpha \in \Hgroup_i$, $\alpha \not\in \image{h_{i-1,i}}$,
$h_{i,j-1} (\alpha) \not\in \image{h_{i-1,j-1}}$, and
$h_{i,j} (\alpha) \in \image{h_{i-1,j}}$.
Writing $r_i$ for the smallest value such that $K_i = f^{-1}[0,r_i]$,
we define the \emph{persistence} of $\alpha$ as $| r_j - r_i |$.

\begin{figure}[bht]
  \centering \resizebox{!}{1.4in}{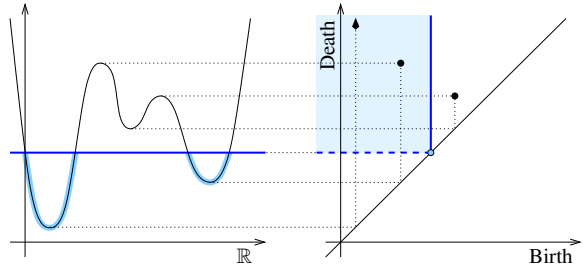}
  \caption{\emph{Left:} the graph of a function on $\Rspace$.
    \emph{Right:} the persistence diagram of the function,
    with two finite points and one point at infinity.
    The components of the highlighted sublevel set are counted
    by the two points in the corresponding upper-left quadrant.}
  \label{fig:persistence}
\end{figure}
To summarize the information in the persistence module, we represent
a coset of homology classes born at $K_i$ and dying entering $K_j$
by the point $(r_i, r_j)$; see Figure \ref{fig:persistence},
setting $r_j = \infty$ if the classes never die.
The result is a multiset of points in the extended plane,
which we refer to as the \emph{persistence diagram} of $f$,
denoted $\Dgm{f}$.
The ranks of the persistent homology groups can be read as the
numbers of points in the upper-left quadrants defined by
corners on or above the diagonal; see again Figure \ref{fig:persistence}.

\paragraph{Stability.}
An important property of persistence diagrams is their stability with respect
to perturbations.
To explain this, we define the \emph{bottleneck distance} between two
persistence diagrams as the length of the longest edge in a minimizing matching.
To finesse the difficulty caused by different cardinalities, we add
infinitely many copies of every point along the diagonal to each diagram
and define
$\bottleneck{\Dgm{f}}{\Dgm{g}} = \inf_\beta \sup_A  \Maxdist{A}{\beta (A)}$,
in which $\beta \colon \Dgm{f} \to \Dgm{g}$ is a bijection
and $A$ is a point in $\Dgm{f}$.
The original proof of stability in \cite{CEH07} bounds the bottleneck
distance by the $L_\infty$-distance between the functions.
In later developments, \cite{BaLe15,CdSGO12} compare the diagrams
directly with the persistence modules they summarize.
Letting $\cal F$ and $\cal G$ be the persistence modules defined by
the sublevel sets of $f$ and $g$,
indexed by real numbers for convenience,
we say they are \emph{$\ee$-interleaved}
if there are maps $\phi_r \colon \Fgroup_r \to \Ggroup_{r+\ee}$
and $\psi_r \colon \Ggroup_r \to \Fgroup_{r+\ee}$ that commute with the
maps within the modules.
The \emph{interleaving distance}, denoted $\interleave{\cal F}{\cal G}$,
is the infimum $\ee \geq 0$ for which $\cal F$ and $\cal G$ are interleaved.
With this notation, we have
\begin{align}
  \bottleneck{\Dgm{f}}{\Dgm{g}} &= \interleave{\cal F}{\cal G}
                                \leq  \Maxdist{f}{g} .
    \label{eqn:stability}
\end{align}
Suppose $\dist$ is a metric, and $X, Y \subseteq \Xspace$
are finite sets with Hausdorff distance $\ee$.
Letting $f$ and $g$ be the restrictions of the half-diameter function
to $\Power{X}$ and to $\Power{Y}$,
it is not difficult to see that the interleaving distance between
the corresponding persistence modules is at most $\ee$.
We state this straightforward consequence of stability
for later reference.
\begin{theorem}[Stability for Metrics]
  \label{thm:StabilityforMetrics}
  Let $\dist \colon \Xspace \times \Xspace \to \Rspace$ be a metric,
  $X, Y \subseteq \Xspace$ finite,
  and $f \colon \Power{X} \to \Rspace$, $g \colon \Power{Y} \to \Rspace$
  induced by the half-diameter function of $\Xspace$ and $\dist$.
  Then the bottleneck distance between $\Dgm{f}$ and $\Dgm{g}$
  is bounded from above by the Hausdorff distance between $X$ and $Y$.
\end{theorem}

\begin{figure}[bht]
  \centering \resizebox{!}{0.40in}{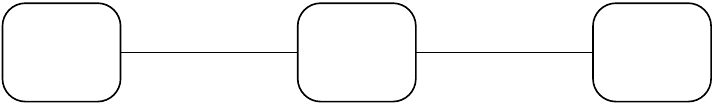}
  \caption{Recall that $\distj = \sqrt{\JS}$, $\disti = \tfrac{1}{2} \Idist{E}$,
    and $\DFunction{\distj}, \DFunction{\disti}, \RFunction{\disti}$
    are the corresponding half-diameter and radius functions.
    Each edge is labeled by two interleaving distances:
    above the edge for $\Xspace = \Rspace_+^n$
    and below the edge for $\Xspace = \Delta$.}
  \label{fig:factors}
\end{figure}

\paragraph{Approximation.}
Since $\sqrt{\JS}$ and $\Idist{E}$ are both metrics,
Theorem \ref{thm:StabilityforMetrics} applies.
Specifically, the mapping from the subsets of $\Xspace$ to the space
of persistence diagrams defined by $\sqrt{\JS}$ is $1$-Lipschitz,
and so is the mapping defined by $\Idist{E}$.
Writing $f_\distj, f_\disti \colon X \to \Rspace$ for the maps
in which $\distj = \sqrt{\JS}$ and $\disti = \tfrac{1}{2} \Idist{E}$,
the persistence diagrams of $f_\distj$ and $f_\disti$ are
generally different, and we can use this difference to compare
the two distances.
By Theorem \ref{thm:JSLvsFI}, we have
\begin{align}
  \tfrac{1}{C} \, \distjFn{x}{y}
    &\leq  \distiFn{x}{y}
     \leq  C \, \distjFn{x}{y} 
    \label{eqn:sevsjs}
\end{align}
for $C^2 = 1 / \ln 2 = 1.442\ldots$,
in which the two distances between $x$ and $y$ are measured in $\Rspace_+^n$.
Rewriting \eqref{eqn:sevsjs} in terms of sublevel sets, we get
$f_\distj^{-1} [0, \tfrac{1}{C} r]  \subseteq  f_\disti^{-1} [0, r]
                                    \subseteq  f_\distj^{-1} [0, C r]$.
To get constant interleaving distance, we re-index the vector spaces
in the persistence modules $\cal F_\distj$ of $f_\distj$
and $\cal F_\disti$ of $f_\disti$ by substituting $\ln r$ for $r$.
Hence, $\interleave{\cal F_\distj}{\cal F_\disti} \leq \ln C = 0.183\ldots$.
By \eqref{eqn:stability}, this implies that the
persistence diagrams of $\cal F_\distj$ and $\cal F_\disti$ are very similar,
with bottleneck distance at most $0.183\ldots$.
We get slightly different results if we measure the distances in $\Delta$.
Specifically, we get $C^2 = \sqrt{2} \pi / (4 \ln 2) = 1.602\ldots$
in \eqref{eqn:sevsjs} and interleaving distance at most
$\ln C = 0.235\ldots$.
\begin{theorem}[Approximation]
  \label{thm:Approximation}
  Let $\distj = \sqrt{\JS}$ and $\disti = \tfrac{1}{2} \Idist{E}$,
  $X \subseteq \Xspace$ finite,
  and ${\cal F_\distj}, {\cal F_\disti}$ the persistence modules
  defined by the restrictions of
  $\ln \DFunction{\distj}, \ln \DFunction{\disti}$ to $\Power{X}$.
  Then $\cal F_\distj$ and $\cal F_\disti$ are interleaved,
  with the interleaving distance labeling the first edge
  in Figure \ref{fig:factors}.
\end{theorem}

\noindent {\sc Remark 6:}
The two constants labeling the first edge in Figure \ref{fig:factors}
can be improved by taking advantage of the asymmetry of the inequalities
in \eqref{eqn:JSvsInf-1} and \eqref{eqn:JSvsInf-2}.
Indeed, while being symmetric, \eqref{eqn:sevsjs} is strictly weaker
than \eqref{eqn:JSvsInf-1}.

\paragraph{Radius function by comparison.}
To offer a perspective, we contrast the half-diameter functions with a
different construction,
which we introduce for the Fisher information metric,
$\disti = \tfrac{1}{2} \Idist{E}$.
Given $\Xspace$, the \emph{radius function}
$\RFunction{\disti} \colon \Power{\Xspace} \to \Rspace$ is defined by mapping
every non-empty subset to
\begin{align}
  \RFunction{\disti} (Q)  &=  \inf_{z \in \Xspace} \sup_{x \in Q}
                              \IdistLn{E}{z}{x} ,
    \label{eqn:radius-function}
\end{align}
which we call the \emph{radius} of $Q$.
Again, the radius function is monotonic,
but in contrast to the half-diameter functions,
the simplices of dimension $2$ and higher encode useful information.
To explain, we write $\Ball{}{r}{x}$ for the set of points
$\IdistLn{E}{x}{z} \leq r$.
The radius function signals at which radius the balls defined by $Q$
have a non-empty common intersection: $r \geq \RFunction{\disti} (Q)$
if and only if $\bigcap_{x \in Q} \Ball{}{r}{x} \neq \emptyset$. 

To see why this property is useful, consider a finite set
$X \subseteq \Xspace$, write $g \colon \Power{X} \to \Rspace$
for the restriction of the radius function,
and note that $G_r = g^{-1} [0,r]$ is a subcomplex of $G = \Power{X}$.
We refer to $G_r$ as the \emph{\v{C}ech complex} of $X$ and $\disti$
for radius $r$.
Furthermore, write $X_r = \bigcup_{x \in X} \Ball{}{r}{x}$ for the union
of the balls with radius $r$.
We thus have two filtrations:
the complexes $G_r$ and the subspaces $X_r$ of $\Xspace$.
The Nerve Theorem implies that $G_r$ and $X_r$ have isomorphic homology groups
\cite{Bor48,Ler45}, and this property extends:
\begin{theorem}[Isomorphism of Modules]
  \label{thm:IsomorphismofModules}
  The persistence modules defined by the complexes $G_r = g^{-1} [0,r]$
  and the subspaces $X_r = \bigcup_{x \in X} \Ball{}{r}{x}$
  are isomorphic and the corresponding persistence diagrams are equal.
\end{theorem}

Theorem \ref{thm:IsomorphismofModules} suggests we consider the common
persistence diagram of the subspaces $X_r$ and the \v{C}ech complexes $G_r$
the "correct" diagram of the data set, $X$.
Not surprisingly, the persistence diagram of the corresponding
Vietoris--Rips complexes is generally different.
We therefore have the opportunity to quantify the error we tolerate
when we compute persistence for the Vietoris--Rips instead of the
\v{C}ech complexes.
Clearly, $\DFunction{\disti} (Q) \leq \RFunction{\disti} (Q)$,
but how different can the two values be?
Considering first the case $\Xspace = \Rspace_+^n$,
we recall Theorem \ref{thm:FItoED} \eqref{eqn:Edistance2}.
As proved in \cite[page 62]{EdHa10}, the radius of a simplex in Euclidean
space is at most $\sqrt{2}$ times the half-length of its longest edge.
The isometry between the Fisher information distance and the Euclidean distance
implies $\RFunction{\disti} (Q) \leq \sqrt{2} \DFunction{\disti} (Q)$
in $\Rspace_+^n$ equipped with the Fisher information metric.
Hence, $\tfrac{1}{C} \DFunction{\disti}(Q) \leq \RFunction{\disti}(Q)
                                           \leq C \DFunction{\disti}(Q)$
with $C^2 = 2$.
The corresponding interleaving distance is
$\ln C = 0.346\ldots$, as recorded in Figure \ref{fig:factors}.
The calculation of the interleaving distance for points in $\Delta$
is similar, except that we do not have a Euclidean result ready to use.
We will prove shortly that for simplices $Q \subseteq \Sspace_+^{n-1}$,
we have
$\tfrac{1}{C} \DFunction{\disti}(Q) \leq \RFunction{\disti}(Q)
                                    \leq C \DFunction{\disti}(Q)$
with $C^2 = 4$.
The corresponding interleaving distance
is $\ln C = 0.693\ldots$.
We thus observe that there is a higher penalty for using Vietoris--Rips
complexes in $\Delta$ than there is in $\Rspace_+^n$.
We finally prove the required Euclidean result.
We recall that the radius and half-diameter of a set $Q$ are defined
in \eqref{eqn:halfdiameter-function} and \eqref{eqn:radius-function}.
\begin{lemma}[Interleaving on the Sphere]
  \label{lem:InterleavingontheSphere}
  Let $Q$ be a finite set of points in $\Sspace_+^{n-1}$.
  Measuring distances in the $(n-1)$-sphere, the radius of $Q$
  is at most twice the half-diameter of $Q$, and this bound is tight
  as $n$ goes to infinity.
\end{lemma}
\myproof
  Suppose first that $Q$ is a regular $(n-1)$-simplex.
  It cannot be larger than the $(n-1)$-simplex $Q_0$ that
  covers the entire positive orthant of the sphere.
  Its vertices are the $n$ unit coordinate vectors,
  and its center lies on the diagonal,
  with all coordinates equal to $1 / \sqrt{n}$.
  The scalar product of the vectors connecting the origin to
  the center and a vertex of $Q_0$ is $1 / \sqrt{n}$,
  which goes to $0$ as $n$ goes to infinity.
\Skip{Drawing $Q_0$ in $\Rspace^n$, we get the standard $(n-1)$-simplex
 whose barycenter is the point
 $\mu$ with coordinates $1/n$.
 Centrally projecting $\mu$ to $\Sspace_+^{n-1}$,
 we get $\varphi (\mu)$ with coordinates $1 / \sqrt{n}$.
 The squared Euclidean distance between the two points is
 \begin{align}
   \Edist{\mu}{\varphi(\mu)}^2
     &=  n \left( \tfrac{1}{\sqrt{n}} - \tfrac{1}{n} \right)^2 
      =  \left( 1 - \tfrac{1}{\sqrt{n}} \right)^2 .
 \end{align}
 In the limit, as $n$ goes to infinity, this distance is $1$.}
  It follows that the spherical distance between the two points
  approaches $\tfrac{\pi}{2}$.
  Since the half-diameter of $Q_0$ is $\tfrac{\pi}{4}$
  and the radius of $Q_0$
  approaches $\tfrac{\pi}{2}$ as $n$ goes to infinity,
  the claimed inequality holds for regular $(n-1)$-simplices
  and cannot be improved
  unless we assume a fixed finite dimension.

  To see that the claimed inequality holds for general sets $Q$,
  we recall that the spherical distance is a metric.
  Let $\Ball{}{r}{x}$ be the ball with Fisher information radius $r$
  centered at $x \in Q$.
  Setting $r$ equal to the diameter of $Q$, every such ball contains all
  points of $Q$, which implies that the common intersection is
  non-empty.
  Hence, the radius is at most the diameter.
\myeop

\section{Discussion}
\label{sec6}

The main contribution of this paper is a connection
between information theory and topological data analysis,
thus complementing the established field of information geometry
\cite{AmNa00} with topological methods.
Concretely, we study notions of distance with information
theoretic foundations, focusing on properties relevant in their
application to topological data analysis.

\noindent The biggest immediate practical gains are the following:
\begin{itemize}\denselist
  \item Computational topology tools working with Euclidean distance can be used for data measured with the 
  Fisher metric. This provides an easy way to experiment with data embedded in information spaces without the need to develop specialized software.
  \item Another option is to use efficient software for the Rips construction like Ripser~\cite{Ripser16}, along with the Jensen-Shannon divergence. 
  This is currently significantly more efficient than using the nerve construction for the entropy balls.
\end{itemize}

\noindent The work reported in this paper raises a number of open questions.
\begin{itemize}\denselist  
  \item The cosine dissimilarity is an effective measure for text documents 
	in the vector space model~\cite{Hua08}. 
    It represents each document as a discrete probability distribution,
    and --- just like the Antonelli isometry --- it maps this distribution to a point
    in the positive orthant of a sphere.
    Overall this computation is equivalent to the Fisher information distance in $\Delta$, except 
	the latter uses the angle and the former, its cosine. 
	This suggests an information-theoretic interpretation for the cosine measure. 
    Using the generalized Antonelli isometry, we can devise similarly simple
    measures for other decomposable Bregman divergences.
    Will they prove equally effective in practice?
    
    \item Theorem \ref{thm:NonconvexDivergenceBalls} presented in Appendix~\ref{app:burgdiv}
    gives $\ln 2 - \tfrac{1}{2}$ as a lower bound and $2 \ln 2 - 1$
    as an upper bound for the squared radius at which the
    Itakura--Saito balls switch from convex to nonconvex.
    The authors of this paper have a proof that the upper bound is
    tight in $\Rspace_+^2$.
    Is this also true in three and higher dimensions?
\end{itemize}

The first concrete application of the theory set up here
is described in~\cite{EdNi16}, where Euclidean and Fisher information point processes
are studied and compared through the lens of integral geometry.
Many challenges remain before topology
can be fully utilized in high-dimensional data analysis tasks.
In particular, a proof of stability of topological descriptors
based on Bregman divergences remains elusive, and we hope that the results
presented here are a step in this direction.

\clearpage

\clearpage \appendix
\section{Burg Entropy by Comparison}
\label{app:burgdiv}

To offer a perspective, we contrast Theorem \ref{thm:ConvexEntropyBalls}
with the analysis of balls defined for a Legendre type function
different from the Shannon entropy.
The \emph{Burg entropy} is the function $G \colon \Rspace_+^n \to \Rspace$
defined by $G(x) = \sum_{i=1}^n [1 - \ln x_i]$.
The corresponding divergence,
\begin{align}
  \Bdist{G}{x}{y}  &=  G(x) - [G(y) + \scalprod{\nabla G(y)}{x-y}]
    =  \sum\nolimits_{i=1}^n [ \ln \tfrac{y_i}{x_i} + \tfrac{x_i}{y_i} - 1 ] ,
\end{align}
is known as the \emph{Itakura--Saito divergence} \cite{ItSa68}.
We note that the Burg entropy is strongly decomposable, which implies
that every one of its divergence functions is strongly decomposable.
By Lemma \ref{lem:ComposableConvexity},
these functions are convex if and only if their components are convex.
Furthermore, the divergence functions of different points can be obtained
by stretching the domain along coordinate directions.
To explain this, let $\One = (1, 1, \ldots, 1) \in \Rspace_+^n$
and write $y/x = (y_1/x_1, y_2/x_2, \ldots, y_n/x_n)$.
\begin{lemma}[Stretch Deformation]
  \label{lem:StretchDeformation}
  Let  $G \colon \Rspace_+^n \to \Rspace$ be the Burg entropy,
  and $x \in \Rspace_+^n$.
  Then the divergence function of $x$ satisfies
  $\DivF{x} (y) = \DivF{\One} (y/x)$.
\end{lemma}
\myproof
  Note that $x/x = \One$.
  Hence, $\DivF{\One} (y/x)$ is equal to
  $\Bdist{G}{x/x}{y/x}       
      =  \sum\nolimits_{i=1}^n [ \ln \tfrac{y_i}{x_i} + \tfrac{x_i}{y_i} - 1 ]$,
  which is $\DivF{x} (y) = \Bdist{G}{x}{y}$.
\myeop

Scaling the components does not affect the convexity or nonconvexity
of the sublevel sets.
Hence, $\Ball{G}{r}{x}$ is convex if and only if $\Ball{G}{r}{\One}$ is convex.
We use this insight to show that under the Burg entropy,
all balls beyond a certain size are nonconvex.
\begin{theorem}[Nonconvex Divergence Balls]
  \label{thm:NonconvexDivergenceBalls}
  Let $G \colon \Rspace_+^n \to \Rspace$ be the Burg entropy in
  $n \geq 2$ dimensions.
  Then a ball $\Ball{G}{r}{x}$ of the Itakura--Saito divergence
  is convex if $r^2 \leq \ln 2 - \tfrac{1}{2}$
  and nonconvex if $r^2 > 2 \ln 2 - 1$.
\end{theorem}
\myproof
  Writing $g(t) = 1 - \ln t$ for the component function of the Burg entropy,
  the component function of the divergence function of $\One \in \Rspace_+^n$ is
  $d(t) =  g(1) - [g(t) + g'(t) (1-t)] =  \ln t + \tfrac{1}{t} - 1$.
  Its first and second derivatives are
  $d'(t)  = \tfrac{1}{t} - \tfrac{1}{t^2}  =  \tfrac{t-1}{t^2}$ and
  $d''(t) =  - \tfrac{1}{t^2} + \tfrac{2}{t^3}  =  \tfrac{-t+2}{t^3}$.
  Hence, $d$ is convex for $t \leq 2$ and concave for $t \geq 2$.
  This implies that $d^{-1} [0,r^2]$
  is convex as long as $r^2 \leq d(2) = \ln 2 - \tfrac{1}{2}$.
  By Lemma \ref{lem:ComposableConvexity}, $\Ball{G}{r}{\One}$
  is convex for $r^2 \leq \ln 2 - \tfrac{1}{2} = 0.193\ldots$.
  To prove the upper bound, we note that Remark 3 applies here.
  Setting $x = (a,b,1,\ldots,1)$ and $y = (b,a,1,\ldots,1)$,
  with $a = 2 + \ee$, $b = 2 + 2\ee$, and $\ee > 0$,
  we get a nonconvex ball of squared radius
  $r^2 = \DivF{\One} (x) = \DivF{\One} (y) = d(a) + d(b)$.
  Hence, $r^2 = 2 \ln 2 - 1 + f(\ee)$,
  in which $f(\ee)$ approaches $0$ as $\ee$ goes to $0$.
  Since $\ee$ can be chosen arbitrarily small, this implies
  that $\Ball{G}{r}{\One}$ is nonconvex for all
  $r^2 > 2 \ln 2 - 1 = 0.386\ldots$.
  By Lemma \ref{lem:StretchDeformation}, the same is true for
  all balls $\Ball{G}{r}{x}$, $x \in \Rspace_+^n$.
\myeop

\Skip{\medskip \noindent {\sc Remark 4:}
  While the balls of the Itakura--Saito divergence can be nonconvex,
  their non-empty common intersections are necessarily contractible,
  as proved in \cite{EdWa16}.
  This implies that the Nerve Theorem applies to any covering with
  such balls.}

\section{Burg Information Metric by Comparison}
\label{app:burginf}

We wish to compare the metrics defined by the Shannon entropy and
the Burg entropy.
Recall that the latter is the Legendre type function
$G \colon \Rspace_+^n \to \Rspace$ defined by mapping $x$ to
$G(x) = \sum_{i=1}^n [1 - \ln x_i]$, which is twice differentiable.
Using its Hessian, we define the corresponding
\emph{Burg information metric},
$\Idist{G} \colon \Rspace_+^n \times \Rspace_+^n \to \Rspace$.
\begin{theorem}[Burg Information to Euclidean Distance]
  \label{thm:BItoED}
  Let $G \colon \Rspace_+^n \to \Rspace$ be the Burg entropy,
  and $x, y \in \Rspace_+^n$.
  Then
  \begin{align}
    \IdistLn{G}{x}{y}  &=  \sqrt{ \tfrac{1}{2} \sum\nolimits_{i=1}^n
                         \left( \ln x_i - \ln y_i \right)^2 } .
      \label{eqn:Gdistance1}
  \end{align}
\end{theorem}
\myproof
  Let $\psi \colon \Rspace_+^n \to \Rspace$ be the diffeomorphism that
  distorts the Burg information metric to the Euclidean metric.
  It decomposes into $n$ copies of $q \colon \Rspace_+ \to \Rspace$.
  By Equation \eqref{eqn:diffeoprime}, the derivative of the
  component function is $q' (t) = 1 / \sqrt{2 t^2}$.
  We get $q (t) = \ln t / \sqrt{2}$ by integration.
  The diffeomorphism maps a point $x$ with coordinates $x_i > 0$
  to the point $\psi (x)$ with coordinates
  $q (x_i) = \ln x_i / \sqrt{2}$, which may be positive, zero, or negative.
  The squared Euclidean distance between $\psi (x)$ and $\psi (y)$ is
  $\tfrac{1}{2} \sum_{i=1}^n (\ln x_i - \ln y_i)^2$,
  which implies \eqref{eqn:Gdistance1}.
\myeop

While the Burg information metric in $\Rspace_+^n$ has a satisfactory description
given in Theorem \ref{thm:BItoED},
the situation is more complicated when we restrict the domain to
$\Delta \subseteq \Rspace_+^n$.
The image of the standard simplex, $\psi (\Delta)$, consists of all points
with coordinates $u_i = q (x_i)$ that satisfy
$e^{\sqrt{2} u_1} + e^{\sqrt{2} u_2} + \ldots + e^{\sqrt{2} u_n}  =  1$.
The thus specified $(n-1)$-dimensional surface
is the graph of a smooth and strictly concave function whose graph
is close to the boundary of the negative orthant.
Denoting this surface by $\Hspace^{n-1}$,
we get an isometry $\psi|_\Delta \colon \Delta \to \Hspace^{n-1}$.

\begin{figure}[bht]
  \vspace{-0.1in}
  \centering \resizebox{!}{2.2in}{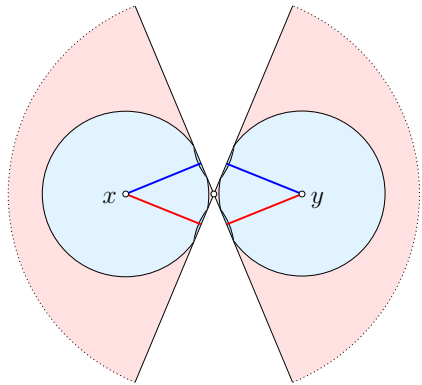}
  \vspace{-0.1in}
  \caption{An unfolding of the cone that is the boundary
    of the negative orthant in $\Rspace^3$.
    The two disks contain both shortest connecting paths and
   form a hole that contains the apex.}
  \label{fig:cone}
\end{figure}
It is not difficult to show that the shortest path between two points
in $\Delta$ is not unique.
We illustrate this by approximating $\Hspace^2$ with the cone that has
an angle $\tfrac{3 \pi}{2}$ at the apex; see Figure \ref{fig:cone}.
Indeed, this cone is isometric to the boundary of the negative orthant.
To draw the cone in the plane, we cut it symmetrically along two half-lines.
Let $x$ and $y$ be opposite from each other at unit distance from the apex.
There are two shortest paths connecting them, by-passing the apex on
different sides.
The disks of radius $\cos \tfrac{\pi}{16}$ centered at $x$
and at $y$ cover both paths but not the apex.
Indeed, the union of the two disks is an annulus with a hole
containing the apex.
To create this hole, the two disks intersect in two disconnected pieces.
The configuration depicted in Figure \ref{fig:cone}
can be approximated arbitrarily closely for $\Hspace^2$,
namely if we increase the radii of the two disks as needed.
In other words, the balls in the Burg information metric violate the
essential assumption of the Nerve Theorem \cite{Bor48,Ler45},
namely the contractibility of non-empty common intersections,
and can therefore not be convex in $\Delta$.

\section{Burbea--Rao Divergence by Comparison}
\label{app:br}

The Jensen--Shannon divergence and the Fisher information metric both generalize
to the case of the Burg entropy, but not all results do.
In particular, the left inequality in \eqref{eqn:JSvsInf-1} extends,
while there is not counterpart to the right inequality.
We call the generalization of the Jensen--Shannon divergence
to Legendre type functions the \emph{Burbea--Rao divergence}.
For the Burg entropy and points $x, y \in \Rspace_+^n$, it is
\begin{align}
  \!\!\BRLs{G}{x}{y}  &=  \tfrac{1}{2} [ \Bdist{G}{x}{\mu}\!+\!\Bdist{G}{y}{\mu} ]
                   =  \tfrac{1}{2} [G(x)\!+\!G(y)] - G(\mu)                  
                   =  \tfrac{1}{2} \sum_{i=1}^n
                      \left[ \ln \tfrac{(x_i+y_i)^2}{4 x_i y_i} \right] ,
    \label{eqn:BR-3}
\end{align}
in which $\mu = \tfrac{x+y}{2}$.
The squared distance between $x$ and $y$ according to the Burg information metric
is given in \eqref{eqn:Gdistance1} as
\begin{align}
  \IdistSq{G}{x}{y}  &=  \tfrac{1}{2} \sum\nolimits_{i=1}^n
                         [ \ln x_i - \ln y_i ]^2 .
    \label{eqn:Bmetric}
\end{align}
\begin{theorem}[Burbea--Rao Divergence vs.\ Burg Information]
  \label{thm:BRLvsBI}
  The Burbea--Rao divergence and the squared Burg information metric satisfy
  $4 \, \BRLs{G}{x}{y}  \leq  \IdistSq{G}{x}{y}$
  for all $x, y \in \Rspace_+^n$.
  Furthermore, for any $C > 0$, there are points $x, y \in \Rspace_+^n$
  for which $\IdistSq{G}{x}{y} > C \, \BRLs{G}{x}{y}$.
\end{theorem}

\myproof
  Both the Burbea--Rao divergence and the squared Burg information metric are decomposable,
  so it suffices to consider the $1$-dimensional case in which we
  write $a$ for $x$ and $b$ for $y$.
  From \eqref{eqn:BR-3} we get
  $\BRLs{G}{Ca}{Cb} = \BRLs{G}{a}{b}$,
  and from \eqref{eqn:Bmetric} we get
  $\IdistSq{G}{Ca}{Cb} = \IdistSq{G}{a}{b}$, for any $C > 0$.
  Setting $b = ta$ and $C = \tfrac{1}{a}$,
  we thus get $\BRLs{G}{a}{b} = \BRLs{G}{1}{t}$
  and $\IdistSq{G}{a}{b} = \IdistSq{G}{1}{t}$.
  We will show that the ratio of the two functions decreases monotonically
  within $1 < t < \infty$,
  attaining its extreme values in the two limits.
  Setting $g(t) = \BRLs{G}{1}{t} / \IdistSq{G}{1}{t}$, we have
  $g(t) = \tfrac{2 \ln (t+1) - \ln (4t)}{( \ln t )^2}$.
  As $t$ goes to infinity, the denominator grows faster than the numerator,
  which implies $\lim_{t \to \infty} g(t) = 0$.
  In other words, for every $C > 0$, there is a sufficiently large
  $t$ such that $\IdistSq{G}{1}{t} > C \, \BRLs{G}{1}{t}$, as claimed.
  At $t = 1$, we get $g(1) = \tfrac{0}{0}$.
  To resolve this ambiguity, we write $g(t) = \tfrac{u(t)}{v(t)}$
  and use the l'Hopital rule by differentiating twice:
  \begin{align}
    u'(t)   &=  \tfrac{t-1}{t (t+1)} ,
    ~~~~~ u''(t)   =  \tfrac{-t^2+2t+1}{t^2 (t+1)^2} ,  \\
    v'(t)   &=  \tfrac{2 \ln t}{t} ,
    ~~~~~~~~ v''(t)   =  \tfrac{2 - \ln t}{t^2} .
  \end{align}
  The second derivatives at $t = 1$ are $u''(1) = \tfrac{1}{2}$
  and $v''(1) = 2$, which gives
  $\lim_{t \to 1} g(t) = \tfrac{1}{4}$, as required.
  To prove that $g$ decreases monotonically, from $\tfrac{1}{4}$ at $t = 1$
  to $0$ at $t = \infty$,
  we compute the first derivative:
  \begin{align}
    g'(t)  &=  \tfrac{(t-1) \ln t - 4 (t+1) \ln (t+1) + 2 (t+1) \ln (4t)}
                    {t (t+1) (\ln t)^3} .
  \end{align}
  The denominator is positive so it suffices to show that the numerator
  is negative, for all $t > 1$.
  Writing $w(t)$ for the numerator of $g'(t)$, we have
  $w(1) = - 8 \ln 2 + 4 \ln 4 = 0$.
  The first two derivatives are
  \begin{align}
    w'(t)  &=  3 \ln t - 4 \ln (t+1) + 4 \ln 2 - 1 + \tfrac{1}{t} , \\
    w''(t) &=  - \tfrac{(t-1)^2}{t^2 (t+1)} .
  \end{align}
  We have $w'(1) = 0$ and $w''(t) < 0$ for all $t > 1$,
  which implies $w'(t) < 0$ for all $t > 1$.
  Hence, $g'(t) < 0$ for all $t > 1$, so $g$ is indeed monotonically
  decreasing, as required.
\myeop

\end{document}

%% file: Figs/relent.pdf_t
\begin{picture}(0,0)%
\includegraphics{Figs/relent.pdf}%
\end{picture}%
\setlength{\unitlength}{4144sp}%
\begingroup\makeatletter\ifx\SetFigFont\undefined%
\gdef\SetFigFont#1#2#3#4#5{%
  \reset@font\fontsize{#1}{#2pt}%
  \fontfamily{#3}\fontseries{#4}\fontshape{#5}%
  \selectfont}%
\fi\endgroup%
\begin{picture}(3399,1789)(1789,-4043)
\put(3376,-2581){\makebox(0,0)[b]{\smash{{\SetFigFont{9}{10.8}{\rmdefault}{\mddefault}{\updefault}{\color[rgb]{0,0,0}$y$}%
}}}}
\put(2341,-2626){\makebox(0,0)[b]{\smash{{\SetFigFont{9}{10.8}{\rmdefault}{\mddefault}{\updefault}{\color[rgb]{0,0,0}$x$}%
}}}}
\put(2251,-3706){\rotatebox{90.0}{\makebox(0,0)[b]{\smash{{\SetFigFont{8}{9.6}{\rmdefault}{\mddefault}{\updefault}{\color[rgb]{0,0,0}$\Bdist{E}{x}{y}$}%
}}}}}
\end{picture}%

%% file: Figs/persistence.pdf_t
\begin{picture}(0,0)%
\includegraphics{Figs/persistence.pdf}%
\end{picture}%
\setlength{\unitlength}{4144sp}%
\begingroup\makeatletter\ifx\SetFigFont\undefined%
\gdef\SetFigFont#1#2#3#4#5{%
  \reset@font\fontsize{#1}{#2pt}%
  \fontfamily{#3}\fontseries{#4}\fontshape{#5}%
  \selectfont}%
\fi\endgroup%
\begin{picture}(8605,4046)(1543,-4085)
\put(5086,-3976){\makebox(0,0)[b]{\smash{{\SetFigFont{20}{24.0}{\rmdefault}{\mddefault}{\updefault}{\color[rgb]{0,0,0}$\Rspace$}%
}}}}
\end{picture}%

%% file: Figs/factors.pdf_t
\begin{picture}(0,0)%
\includegraphics{Figs/factors.pdf}%
\end{picture}%
\setlength{\unitlength}{4144sp}%
\begingroup\makeatletter\ifx\SetFigFont\undefined%
\gdef\SetFigFont#1#2#3#4#5{%
  \reset@font\fontsize{#1}{#2pt}%
  \fontfamily{#3}\fontseries{#4}\fontshape{#5}%
  \selectfont}%
\fi\endgroup%
\begin{picture}(3264,474)(4039,-3943)
\put(4996,-3661){\makebox(0,0)[b]{\smash{{\SetFigFont{8}{9.6}{\rmdefault}{\mddefault}{\updefault}{\color[rgb]{0,0,0}$0.183\ldots$}%
}}}}
\put(6346,-3661){\makebox(0,0)[b]{\smash{{\SetFigFont{8}{9.6}{\rmdefault}{\mddefault}{\updefault}{\color[rgb]{0,0,0}$0.346\ldots$}%
}}}}
\put(4996,-3841){\makebox(0,0)[b]{\smash{{\SetFigFont{8}{9.6}{\rmdefault}{\mddefault}{\updefault}{\color[rgb]{0,0,0}$0.235\ldots$}%
}}}}
\put(6346,-3841){\makebox(0,0)[b]{\smash{{\SetFigFont{8}{9.6}{\rmdefault}{\mddefault}{\updefault}{\color[rgb]{0,0,0}$0.693\ldots$}%
}}}}
\put(7021,-3751){\makebox(0,0)[b]{\smash{{\SetFigFont{14}{16.8}{\rmdefault}{\mddefault}{\updefault}{\color[rgb]{0,0,0}$\RFunction{\disti}$}%
}}}}
\put(5671,-3751){\makebox(0,0)[b]{\smash{{\SetFigFont{14}{16.8}{\rmdefault}{\mddefault}{\updefault}{\color[rgb]{0,0,0}$\DFunction{\disti}$}%
}}}}
\put(4321,-3751){\makebox(0,0)[b]{\smash{{\SetFigFont{14}{16.8}{\rmdefault}{\mddefault}{\updefault}{\color[rgb]{0,0,0}$\DFunction{\distj}$}%
}}}}
\end{picture}%

%% file: Figs/cone.pdf_t
\begin{picture}(0,0)%
\includegraphics{Figs/cone.pdf}%
\end{picture}%
\setlength{\unitlength}{4144sp}%
\begingroup\makeatletter\ifx\SetFigFont\undefined%
\gdef\SetFigFont#1#2#3#4#5{%
  \reset@font\fontsize{#1}{#2pt}%
  \fontfamily{#3}\fontseries{#4}\fontshape{#5}%
  \selectfont}%
\fi\endgroup%
\begin{picture}(6319,6504)(1341,-6913)
\put(3016,-3796){\makebox(0,0)[rb]{\smash{{\SetFigFont{25}{30.0}{\rmdefault}{\mddefault}{\updefault}{\color[rgb]{0,0,0}$x$}%
}}}}
\put(5986,-3796){\makebox(0,0)[lb]{\smash{{\SetFigFont{25}{30.0}{\rmdefault}{\mddefault}{\updefault}{\color[rgb]{0,0,0}$y$}%
}}}}
\end{picture}%